# Thermal Visualization of Buried Interfaces by Transient and Steady-State Responses of Time-Domain Thermoreflectance

Zhe Cheng[1], Fengwen Mu[2,*], Xiaoyang Ji[1], Tiangui You[3], Wenhui Xu[3], Tadatomo Suga[4], Xin Ou[3], David G. Cahill[1], Samuel Graham[5,*]

[1] Department of Materials Science and Engineering and Materials Research Laboratory, University of Illinois at Urbana-Champaign, Urbana, IL 61801, USA.

[2] Kagami Memorial Research Institute for Materials Science and Technology, Waseda University, Shinjuku, Tokyo 169-0051, Japan

[3] State Key Laboratory of Functional Materials for Informatics, Shanghai Institute of Microsystem and Information Technology, Chinese Academy of Sciences, 200050, China

[4] Collaborative Research Center, Meisei University, Hino-shi, Tokyo 191-8506, Japan

[5] George W. Woodruff School of Mechanical Engineering, Georgia Institute of Technology, Atlanta, Georgia 30332, USA

[*]Corresponding authors: mufengwen123@gmail.com; sgraham@gatech.edu

# ABSTRACT

Thermal resistances from interfaces impede heat dissipation in micro/nanoscale electronics, especially for high-power electronics. Despite the growing importance of understanding interfacial thermal transport, advanced thermal characterization techniques which can visualize thermal conductance across buried interfaces, especially for nonmetal-nonmetal interfaces, are still under development. This work reports a dual-modulation-frequency TDTR mapping technique to visualize the thermal conduction across buried semiconductor interfaces for β-$Ga_2O_3$-SiC samples. Both the β-$Ga_2O_3$ thermal conductivity and the buried β-$Ga_2O_3$-SiC thermal boundary conductance (TBC) are visualized for an area of 200 μm x 200 μm. Areas with low TBC values (≤20 MW/$m^2$-K) are successfully identified on the TBC map, which correspond to weakly bonded interfaces caused by high-temperature annealing. The steady-state temperature rise (detector voltage), usually ignored in TDTR measurements, is found to be able to probe TBC variations of the buried interfaces without the limit of thermal penetration depth. This technique can be applied to detect defects/voids in deeply buried heterogeneous interfaces non-destructively, and also opens a door for the visualization of thermal conductance in nanoscale nonhomogeneous structures.

# INTRODUCTION

With the minimization of microelectronics, and the increasing power and frequency of power and radio-frequency (RF) electronics, proper thermal management in these devices are essential for stable and reliable performance.[1] Thermal resistances derived from interfaces in these systems show growing importance and account for a large even dominant part of the total thermal resistances.[2-4] However, advanced thermal characterization techniques to probe thermal conduction in heterogeneous or nonhomogeneous structures are under development.[5-12] Ultrafast laser based pump-probe techniques, such as time-domain thermoreflectance (TDTR) and frequency-domain thermoreflectance (FDTR), have been used to map nonhomogeneous samples and study the relation between local thermal conductivity variation and microscale structural features such as grain boundaries, composition variations, and interface quality.[5,6,8,9] To perform TDTR or FDTR measurements, a metal transducer layer is usually coated on the sample surface. The thermal boundary conductance (TBC) of the metal transducer-substrate interfaces can be easily measured with high spatial resolution.[7,10,11]

For nonmetal-nonmetal interfaces, the mapping of thermal conductance has not been reported previously partly due to the small sensitivity resulting from the limited thermal penetration depth and the difficulty to separate multiple unknown parameters, despite the great importance of understanding the nonhomogeneous thermal transport across these heterogeneous interfaces, for both fundamental thermal science and defect detection in semiconductor devices.[12] TDTR measures the transient thermal response of laser heating on the sample surface by periodically heating the sample surface (pump beam) and monitoring its temperature variation via thermoreflectance (probe beam).[13] By taking advantage of the transient response, the measurements exclude the possible errors from the variations from laser power and thermoreflectance coefficient of the transducer. However, the modulation frequency of the pump beam limits the thermal penetration depth. Only the part of the sample near the sample surface can be probed, leading to small sensitivity of measuring deeply buried interfaces.

In this work, we use a double-frequency TDTR mapping technique to visualize the thermal conductance across buried β-$Ga_2O_3$-SiC interfaces. A measurement with high modulation frequency is performed to probe the thermal conductivity of the monocrystalline β-$Ga_2O_3$ layer while a measurement with low modulation frequency is performed to measure the β-$Ga_2O_3$-SiC TBC. Moreover, we utilize the steady-state heating data of TDTR to infer the structural inhomogeneity of buried interfaces, which is usually ignored in the applications of TDTR.

**RESULTS AND DISCUSSIONS**

Three bonded β-$Ga_2O_3$-SiC samples are studied in this work. The monocrystalline β-$Ga_2O_3$ layers were exfoliated from a bulk crystal by ion-cutting and bonded to SiC by surface activated bonding technique. The details about sample fabrication can be found in the Experimental Section. The three samples are ~300-nm-thick β-$Ga_2O_3$ layers bonded with SiC substrates. Sample_N was annealed in $N_2$ at 800 °C while Sample_O was annealed in $O_2$ at 800 °C. Sample_AB was tested as bonded without annealing. A ~80-nm-thick Al was coated on the sample surface as TDTR transducer. The thermal conductivity of the SiC substrate was measured from the backside of the sample to be 381 W/m-K, which was used as the SiC thermal conductivity in the data fitting of other measurements.

The measured thermal conductivity of the β-Ga$_2$O$_3$ layers in Sample_N, Sample_O, and Sample_AB are shown in Figure 1(a). The two annealed samples have similar thermal conductivities and are more than twice as large as the as-bonded one. The annealing removes the ion-implantation-induced strain and restores the lattice distortion in the β-Ga$_2$O$_3$ layers, resulting in the increase in thermal conductivity, similar to our previous work.[14] The β-Ga$_2$O$_3$ thermal conductivity of Sample_N and Sample_O do not show any modulation-frequency dependence, showing that the β-Ga$_2$O$_3$ defect/vacancy concentrations are relatively uniform. The strong modulation frequency dependence of the measured β-Ga$_2$O$_3$ thermal conductivity of Sample_AB are shown in Figure 1(b). Each modulation frequency corresponds to a thermal penetration depth, as shown in Figure 1(c). The β-Ga$_2$O$_3$ film thickness of Sample_AB is marked as a dash line. When the thermal penetration depth is larger than the film thickness, the laser heating penetrates through the β-Ga$_2$O$_3$ film. The measured β-Ga$_2$O$_3$ thermal conductivity keeps constant. The penetration-depth dependent thermal conductivity shows that Sample_AB has a gradient defect/vacancy concentration distribution, as shown in Figure 1(d). The top part of the β-Ga$_2$O$_3$ layer in Figure 1(d) is close to the exfoliation interface where large amount of implanted hydrogen ions are accumulated to break the β-Ga$_2$O$_3$ thin layer from the β-Ga$_2$O$_3$ bulk crystal.

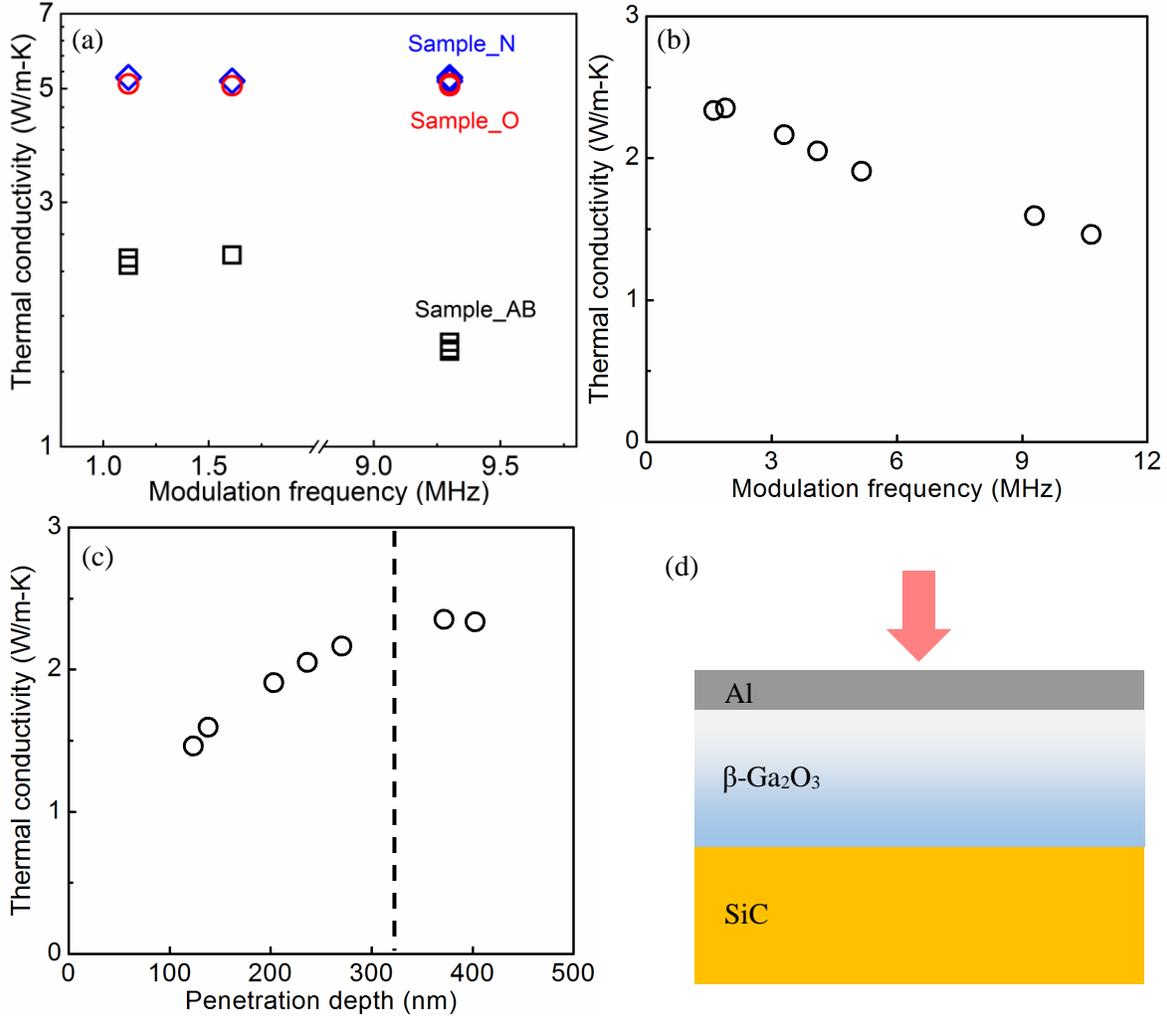

Figure 1. (a) Thermal conductivity comparison of Sample_N, Sample_O, and Sample_AB. (b) Frequency-dependent thermal conductivity of Sample_AB. (c) Penetration-depth-dependent thermal conductivity of Sample_AB. (d) Schematic diagram of the cross section of Sample_AB.

In TDTR measurements, the sensitivity of measuring a specific unknown parameter is defined as

$$S_i = \frac{\partial \ln(-V_{in}/V_{out})}{\partial \ln(p_i)}, \quad (1)$$

where $S_i$ is the sensitivity to parameter $i$, $-V_{in}/V_{out}$ is the TDTR signal, $p_i$ is the value of parameter $i$. For the sample structures in this work, there are three unknown parameters: Al-β-$Ga_2O_3$ TBC, β-$Ga_2O_3$ thermal conductivity ($\kappa$), and β-$Ga_2O_3$-SiC TBC. As shown in Figure 2(a), the calculated sensitivity of each parameter changes with the delay time of TDTR. The dash line shows that the sensitivity of the Al-β-Ga$_2$O$_3$ TBC is zero at a delay time of 650 ps with a modulation frequency of 9.3 MHz on Sample_O. At this delay time, the measured TDTR ratio is not affected by the Al-β-Ga$_2$O$_3$ TBC. The schematic diagram of the transient heating of the sample surface with a modulation frequency of 9.3 MHz is shown in Figure 2(b). The thermal penetration depth is smaller than the film thickness. Therefore, TDTR is sensitive to the β-Ga$_2$O$_3$ $\kappa$ but the sensitivity of the β-Ga$_2$O$_3$-SiC TBC is almost zero. At a delay time of 650 ps, the TDTR ratio is only related to one unknown parameter (β-Ga$_2$O$_3$ $\kappa$).

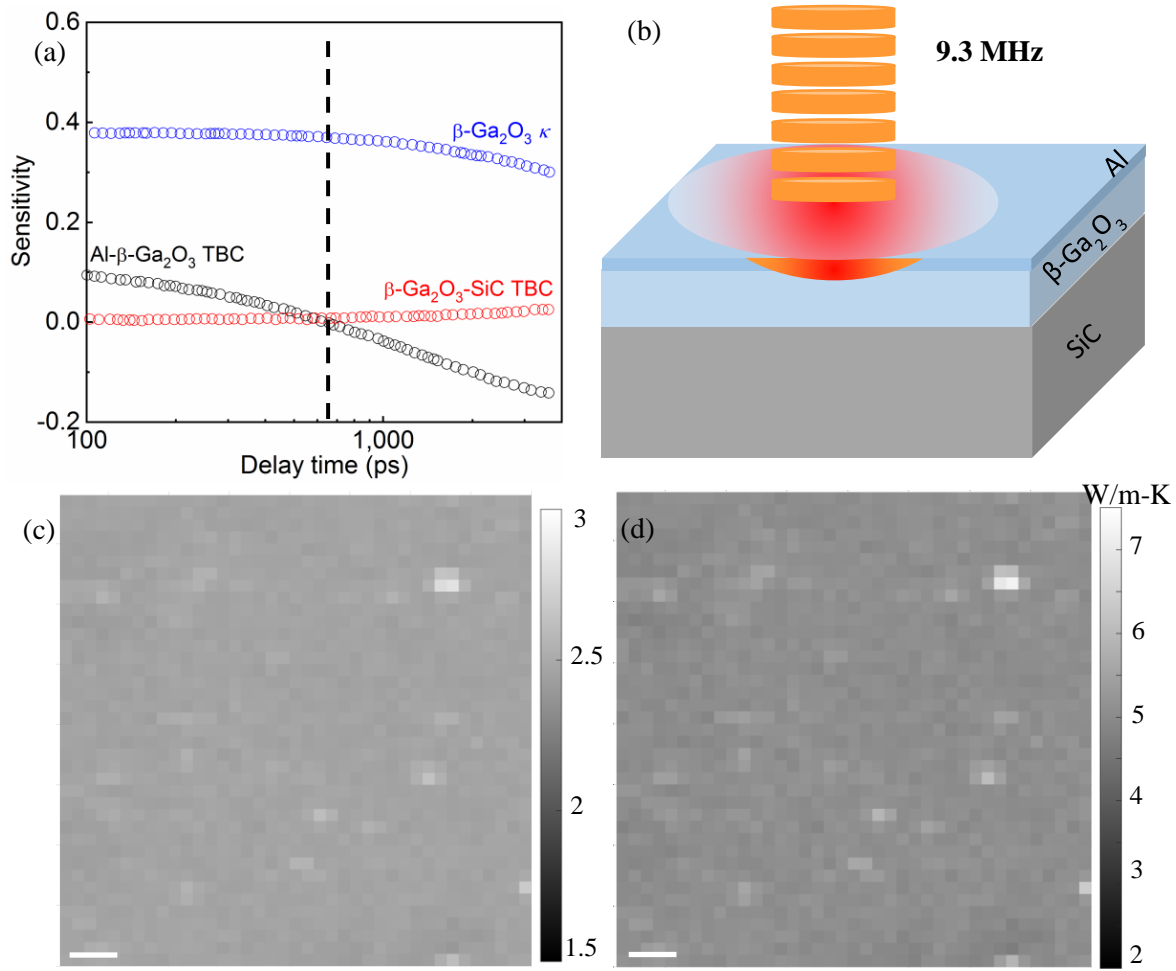

Figure 2. (a) TDTR sensitivity of Al-β-Ga$_2$O$_3$ TBC, β-Ga$_2$O$_3$ $\kappa$, and β-Ga$_2$O$_3$-SiC TBC with a modulation frequency of 9.3 MHz on Sample_O. The dash line shows that the sensitivity of the Al-β-Ga$_2$O$_3$ TBC is zero at delay time of 650 ps. (b) Schematic diagram of TDTR measurements with a modulation frequency of 9.3 MHz. High modulation frequency results in a short thermal penetration depth and low sensitivity of the buried β-Ga$_2$O$_3$-SiC TBC. (c) The TDTR ratio map of Sample_O. The scale bar is 20 μm and the mapped area is 200 μm x 200 μm. (d) The β-Ga$_2$O$_3$ $\kappa$ map of Sample_O. The scale bar is 20 μm.

Figure 2(c) shows a 200 μm x 200 μm map of the TDTR ratio. The step size of the mapping is 5 μm. A 10X objective was used with a laser spot size of 5.5 μm in radius. The dwell time for each step is 0.5 s and the time traveling from the last spot to the first spot on the next line is 1 s. All the mapping parameters mentioned here are the same for the following mappings. The β-Ga$_2$O$_3$ $\kappa$ distribution is shown in Figure 2(d). Most of the spots have a thermal conductivity close to 5 W/m-K, which is close to the measured value of the full-range TDTR measurements in Figure 1(a).

Unlike the high modulation frequency used to measure the β-Ga$_2$O$_3$ $\kappa$ in Figure 2(a), Figure 3(a) shows the sensitivity of Al-β-Ga$_2$O$_3$ TBC, β-Ga$_2$O$_3$ $\kappa$, and β-Ga$_2$O$_3$-SiC TBC with a modulation frequency of 1.61 MHz on Sample_O. The low modulation frequency results in a large thermal penetration depth, as shown in the schematic diagram in Figure 3(b). The thermal penetration depth goes through the β-Ga$_2$O$_3$ layer and TDTR has enough sensitivity to probe the Ga$_2$O$_3$-SiC TBC. At 330 ps as shown by the dash line in Figure 3(a), the sensitivity of the Al-β-Ga$_2$O$_3$ TBC is zero. The measured TDTR ratio is only related to β-Ga$_2$O$_3$ $\kappa$ and β-Ga$_2$O$_3$-SiC TBC. The β-Ga$_2$O$_3$ $\kappa$ at each spot has been measured with the 9.3 MHz modulation frequency as discussed above. The measured TDTR ratio with a modulation frequency of 1.61 MHz is shown in Figure 3(c). It is obvious that there are some locations with very low TDTR ratios. By fitting the TDTR ratio, the β-Ga$_2$O$_3$-SiC TBC map can be extracted with the measured β-Ga$_2$O$_3$ $\kappa$ map, as shown in Figure 3(d). The low TDTR ratio corresponds to a low β-Ga$_2$O$_3$-SiC TBC. The measured β-Ga$_2$O$_3$-SiC TBC is as low as 20 MW/m$^2$-K, which is close to the TBC values of van der Waals bonded interfaces.[15] These areas should have very weak bonds between β-Ga$_2$O$_3$ and SiC substrates. A red dash circle and a blue dash circle are used to mark a typical well-bonded area and a weakly-bonded area. Full-range TDTR scans will be performed on these spots to check the mapping results and presented later in this work. The visualization capability shown here can be applied to detect defects in buried semiconductor structures.

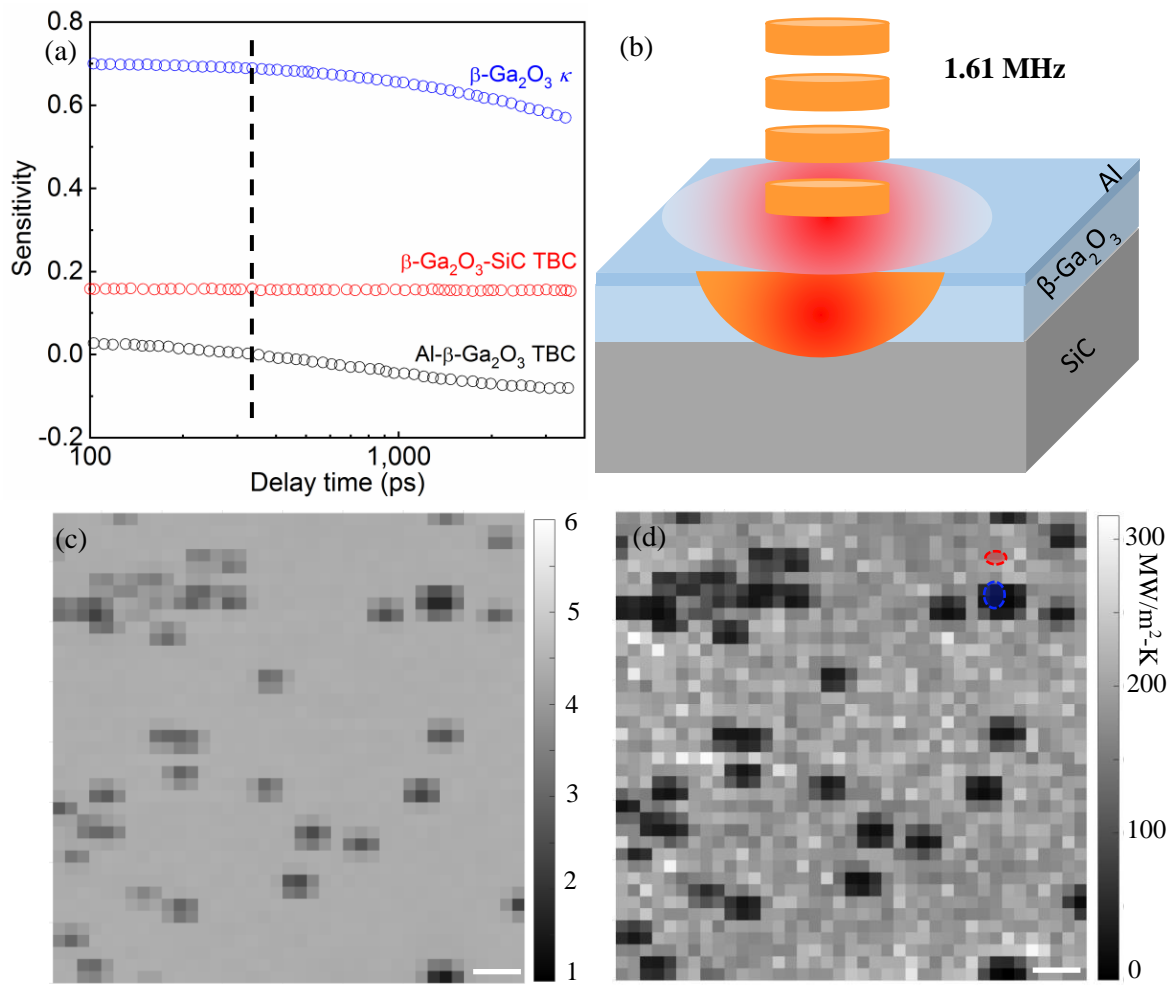

Figure 3. TDTR sensitivity of Al-β-Ga$_2$O$_3$ TBC, β-Ga$_2$O$_3$ $\kappa$, and β-Ga$_2$O$_3$-SiC TBC with a modulation frequency of 1.61 MHz on Sample_O. The dash line shows that the sensitivity of Al-β-Ga$_2$O$_3$ TBC is zero at a delay time of 330 ps. (b) Schematic diagram of TDTR measurements with a modulation frequency of 1.61 MHz. Low modulation frequency results in a large thermal penetration depth and high sensitivity of the buried β-Ga$_2$O$_3$-SiC TBC. (c) The TDTR ratio map of Sample_O. The scale bar is 20 μm and the mapped area is 200 μm x 200 μm. (d) The β-Ga$_2$O$_3$-SiC TBC map of Sample_O. The scale bar is 20 μm. The red and blue dash circles are used to mark a typical well-bonded and weakly bonded spots.

Figure 4(a) shows the statistical distribution of the mapped β-Ga$_2$O$_3$ thermal conductivity. The measured thermal conductivity is quite uniform. More than 75% of the mapped spots have a thermal conductivity between 5 - 5.25 W/m-K. About 95% of the mapped spots have a thermal conductivity between 4.75 - 5.25 W/m-K. This is consistent with the results measured by the full range TDTR scans, as shown in Figure 1(a). Figure 4(b) shows the statistical distribution of the mapped β-Ga$_2$O$_3$-SiC TBC. The measured TBC values spread over a wide range, indicating the non-uniform quality of the buried β-Ga$_2$O$_3$-SiC interfaces. There are a large number of spots which have TBC lower than 50 MW/m$^2$-K. The sensitivity is small (<0.1) for high TBC values (>150 MW/m$^2$-K). Here, we only focus on the areas with low TBC values. These spots are weakly bonded and lead to heat dissipation challenges, which need to be identified in the buried semiconductor interfaces.

To double check our mapped results, a typical well-bonded area and a typical weakly bonded area are chosen to perform full-range TDTR scans, as marked by red and blue dash circles in Figure 3(d). Figure 4(c) shows the acoustic echoes of the Al-β-Ga$_2$O$_3$ interfaces and the β-Ga$_2$O$_3$-SiC interfaces measured by the picosecond acoustic technique on the selected areas. The echoes at a delay time of ~25 ps are from the Al-β-Ga$_2$O$_3$ interfaces while the echoes at a delay time of ~118 ps are from the β-Ga$_2$O$_3$-SiC interfaces. The echoes of the Al-β-Ga$_2$O$_3$ interfaces for all three measurements are the same. However, the echoes from the spots with weakly bonded β-Ga$_2$O$_3$-SiC interfaces are much stronger than the echo of the spot with well-bonded β-Ga$_2$O$_3$-SiC interface. The echoes confirm the structural difference of the β-Ga$_2$O$_3$-SiC interfaces between these spots.

Figure 4(d) shows the full-range TDTR ratios measured on the selected areas. The ratio of the measurement on the well-bonded spot is much larger than those of the weakly bonded spots. This also confirms the difference of thermal properties of these areas. By fitting the full-range TDTR ratios, the fitted β-$Ga_2O_3$-SiC TBC is about 10 MW/$m^2$-K for the weakly bonded β-$Ga_2O_3$-SiC interfaces and about 160 MW/$m^2$-K for the well-bonded β-$Ga_2O_3$-SiC interfaces, which are close to the mapped TBC values.

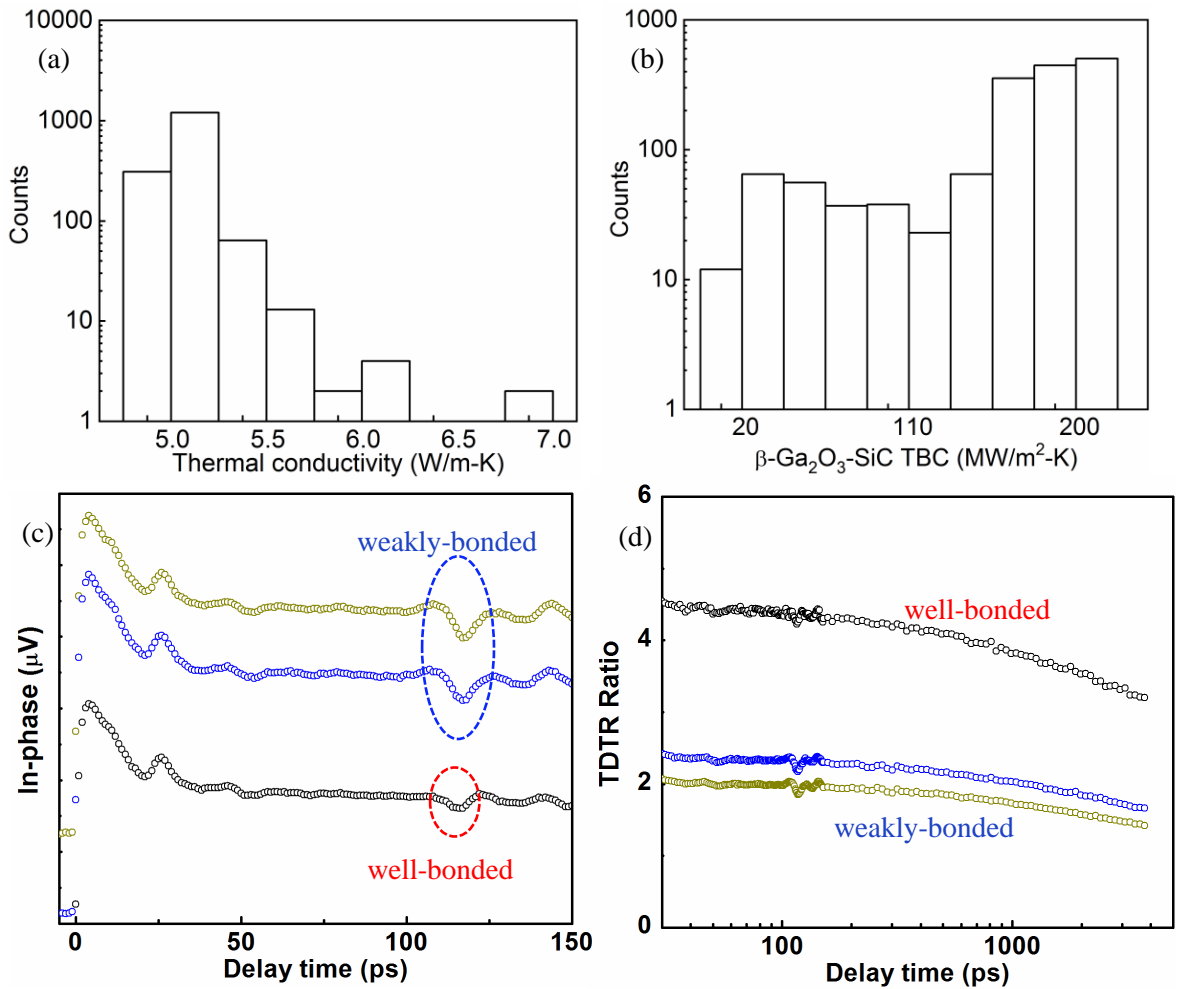

Figure 4. (a) Statistical distribution of the mapped β-$Ga_2O_3$ $\kappa$. (b) Statistical distribution of the mapped β-$Ga_2O_3$-SiC TBC. (c) Acoustic echoes of the Al-β-$Ga_2O_3$ interfaces and the β-$Ga_2O_3$-

SiC interfaces measured by the picosecond acoustic technique on the selected areas as marked by red and blue circles in Figure 3(d). (d) TDTR ratio curves of measurements on the selected areas as marked by red and blue circles in Figure 3(d).

Figures 5(a-b) show the detector voltage of Sample_O with modulation frequencies of 9.3 MHz and 1.61 MHz. The detector voltage measures the power of the reflected probe beam from the sample surface. If we assume the laser power does not change, the variation of the detector voltage during the mapping is proportional to the variation of the reflectivity. The variation of reflectivity is proportional to the variation of surface temperature if the Al quality is uniform. The maps in Figures 5(a-b) show the steady-state temperature rises due to the TDTR laser heating with the laser power decay corrected in the data processing. More details can be found in the Supporting Information. Previous TDTR measurements usually ignore this steady-state heating data and the information which can be extracted, because the reflectivity is also sensitive to local Al quality and is hard to be quantified accurately. However, for well-polished sample surfaces, the Al quality is consistent for the same deposition on the same sample. The relative variation of the surface reflectivity shows the relative variation of the surface temperature on different spots, and corresponding thermal resistances from the sample surface to heat sinks on different spots. The TBC map is also included in Figure 5(c) as comparison. The TBC map matches with the detector voltage maps, even for the 9.3 MHz measurements. These TDTR measurements with high modulation frequency have no sensitivity to the buried interfaces while the steady-state heating captures the information of the buried interfaces. As shown in Figure 5(d), the steady-state nature enables the deep probing into the samples, inferring thermal properties of the buried structures,

not limited by the thermal penetration depth in TDTR measurements. This can be possibly developed to be a tool for nondestructive defects/voids testing in semiconductor devices.

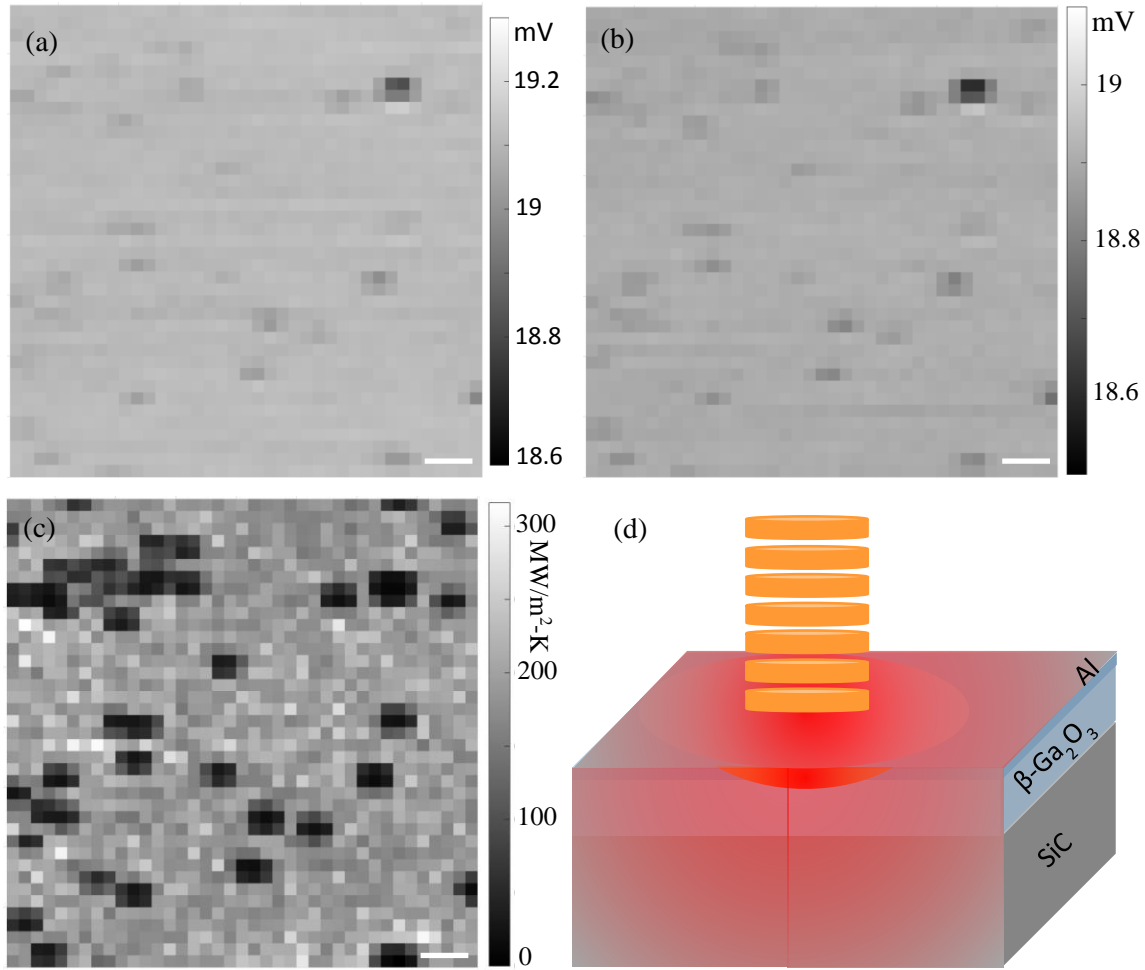

Figure 5. (a) The detector voltage of Sample_O with a modulation frequency of 9.3 MHz. (b) The detector voltage of Sample_O with a modulation frequency of 1.61 MHz. (c) The β-Ga$_2$O$_3$-SiC TBC is included for comparison. (d) Schematic diagram of TDTR measurements with a modulation frequency of 9.3 MHz shows both transient and steady-state heating.

Sample_AB is also mapped with a modulation frequency of 9.3 MHz. Sample_AB has a much lower thermal conductivity than Sample_N and Sample_O, which reduces the thermal penetration depth. TDTR measurements do not have enough sensitivity to the buried β-Ga$_2$O$_3$-SiC TBC even with low modulation frequency. Therefore, the steady-state heating data is used to infer the thermal properties of the interfaces among different spots. As shown in Figure 6(a), the sensitivity of the Al-β-Ga$_2$O$_3$ TBC is zero at the delay time of 580 ps. The sensitivity of the buried β-Ga$_2$O$_3$-SiC TBC is close to zero for all the delay times. Therefore, the TDTR ratio is only related to the β-Ga$_2$O$_3$ $\kappa$. As shown in Figure 6(b) captured by a charge-coupled device (CCD) camera, an area with some defects on the surface is chosen for the mapping, marked by the red dash square. The defect area is marked by the white dash rectangle. The spot in the upper left corner is from the TDTR laser. When the laser beam is focused on the surface, the apparent array of secondary spots marked by red dash circles shows up which is a known artifact of our CCD camera. These spots are not from the reflections of surface defects on the sample.

Figure 6(c) shows the mapped ratio of Sample_AB. The measured TDTR ratio is very uniform except the defective area. The TDTR ratio is fitted to an analytical heat transfer solution of the sample structure to obtain the β-Ga$_2$O$_3$ $\kappa$. The mapped β-Ga$_2$O$_3$ $\kappa$ is shown in Figure 6(d). The uniform $\kappa$ map indicates the uniform quality of the β-Ga$_2$O$_3$ thin films after exfoliation from the bulk β-Ga$_2$O$_3$ crystal, unlike the gradient cross-plane thermal conductivity caused by the gradient defect concentrations.

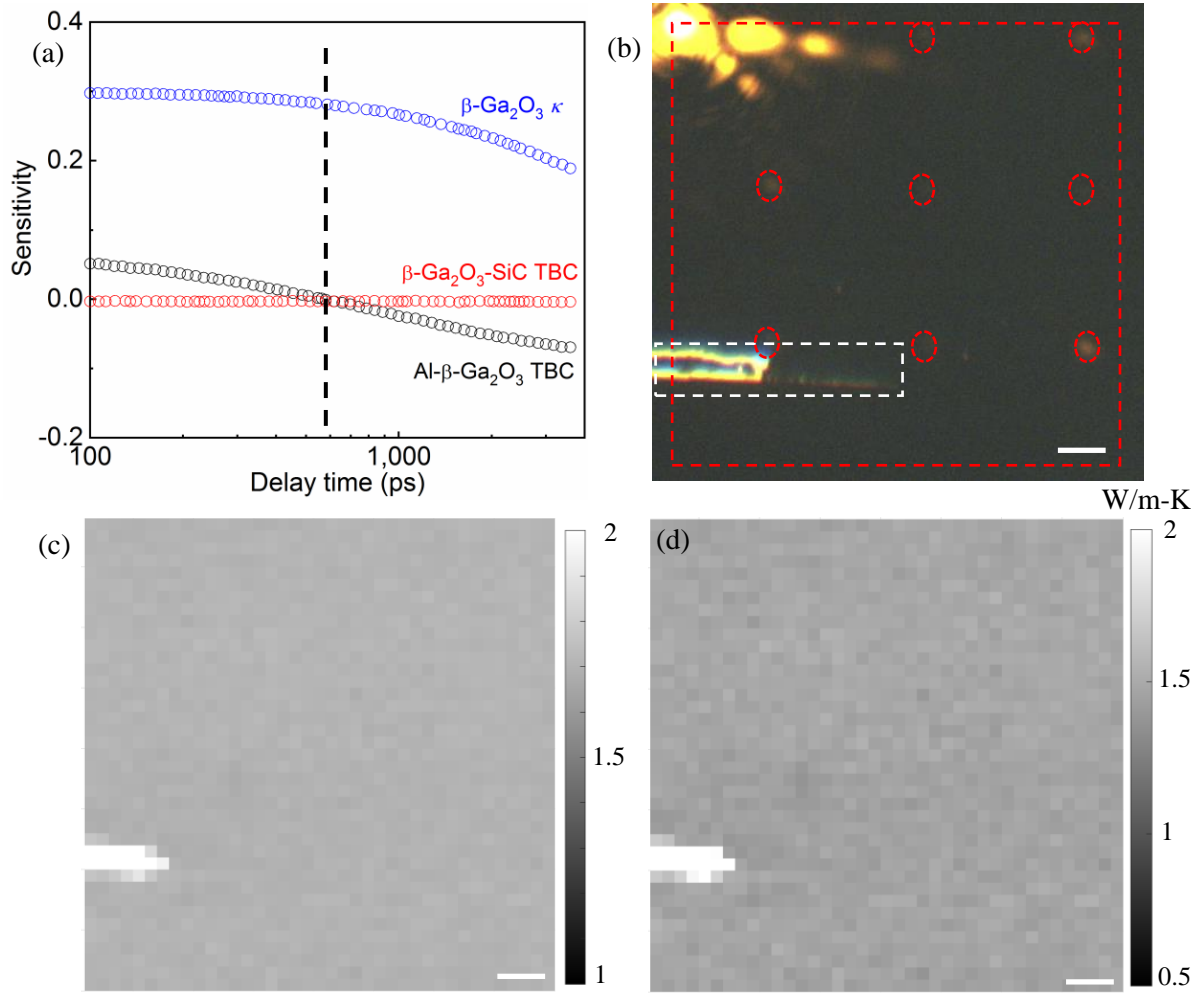

Figure 6. (a) TDTR sensitivity of Al-β-Ga$_2$O$_3$ TBC, β-Ga$_2$O$_3$ $\kappa$, and β-Ga$_2$O$_3$-SiC TBC with a modulation frequency of 9.3 MHz on Sample_AB. The dash line shows that the sensitivity of the Al-β-Ga$_2$O$_3$ TBC is zero at delay time of 580 ps. (b) Sample surface captured by a CCD camera. The red dash line square marks the mapped area. A surface defect is included as a marker in the measurements for comparison. The TDTR laser spot shoots on the upper left corner of the red square. The scale bar is 20 μm and the mapped area is 200 μm x 200 μm. (c) The TDTR ratio map of Sample_AB. The scale bar is 20 μm. (d) The β-Ga$_2$O$_3$ $\kappa$ map of Sample_AB. The scale bar is 20 μm.

The detector voltage of the mapping on Sample_AB with a modulation frequency of 9.3 MHz is shown in Figure 7. The uniform detector voltage distribution shows uniform distributed temperature rises and corresponding uniform distributed thermal resistances, unlike those of Sample_O. According to the above discussion about the steady-state heating of TDTR measurements, we speculate that the buried β-$Ga_2O_3$-SiC TBC distribution is uniform. The difference of sample fabrication between Sample_O and Sample_AB is annealing. Sample_O is annealed at 800 °C in $O_2$ while Sample_AB is as-bonded without annealing. The high temperature annealing improves the quality of the β-$Ga_2O_3$ layer but induces the non-uniformity in the buried interfaces possibly due to the local stress and interfacial diffusion.

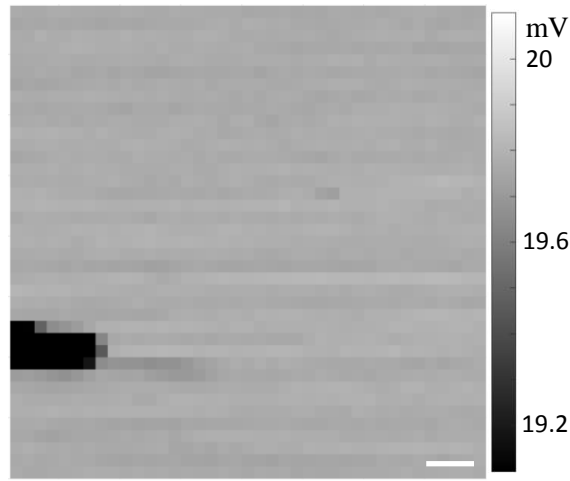

Figure 7. The detector voltage map of Sample_AB with a modulation frequency of 9.3 MHz. The scale bar is 20 μm and the mapped area is 200 μm x 200 μm.

## CONCLUSIONS

This work reports a dual-modulation-frequency TDTR mapping technique to visualize the thermal conductance across buried semiconductor interfaces. The mapped β-$Ga_2O_3$ thermal conductivity is uniform (about 5 W/m-K for a sample annealed at 800 °C in $O_2$). We observe that some areas with very low TBC values ($\leq$20 MW/m$^2$-K) on the TBC map, which correspond to weakly bonded interfaces caused by high-temperature annealing. The steady-state heating data in the TDTR measurements are found to be amenable to probing TBC variation of buried interfaces. The temperature distribution map obtained from the reflectivity map and detector voltage map matches well with the measured TBC map of the buried interfaces. After mapping an as-bonded β-$Ga_2O_3$-SiC sample with a high modulation frequency of 9.3 MHz, both a uniform thermal conductivity map and a uniform detector voltage map are obtained. We speculate that the TBC map of the buried β-$Ga_2O_3$-SiC interfaces are also uniform. Additionally, in the cross-plane direction, the as-bonded β-$Ga_2O_3$ film has a gradient thermal conductivity due to the nonhomogeneous ion implantation concentrations. Our work is able to detect defects/voids in deeply buried heterogeneous interfaces nondestructively, and enables the visualization of thermal conduction in nanoscale nonhomogeneous structures.

## EXPERIMENTAL SECTION

**Sample Preparation.** The 4-inch on-axis (0001) 4H-SiC wafers and the 2-inch ($\bar{2}$01) β-$Ga_2O_3$ wafers were purchased from SICC Co., Ltd and Novel Crystal Technology Inc. The surfaces were polished by chemical mechanical polishing (CMP) to a root-mean-square surface roughness values of ~0.3 nm and ~0.27 nm for SiC and β-$Ga_2O_3$, respectively. The β-$Ga_2O_3$ wafers were implanted by hydrogen ions with an energy of 35 keV and an implantation dose of ~1×10$^{17}$ cm$^{-2}$ at room temperature with a 7° tilt angle. The hydrogen ions accumulate in the β-$Ga_2O_3$ at 200-400 nm beneath the surface. The β-$Ga_2O_3$ is then bonded with the SiC wafer by surface activated bonding at room temperature. The background pressure kept at 5×10$^{-6}$ Pa. A silicon source is added into the Ar source to ion-implant both the SiC and β-Ga$_2$O$_3$ surfaces with a power of 1.0 kV and 100 mA. The chamber pressure is about 0.4 Pa. More details about the bonding and ion implantation can be found in the literature.[14,15] The implanted hydrogen ions induced the exfoliation of a β-Ga$_2$O$_3$ thin film after heating the bonded wafers at 450 °C. The exfoliated β-Ga$_2$O$_3$ film is polished by CMP to a thickness of ~300 nm. Sample_AB is the as-bonded sample while Sample_N and Sample_O are annealed at 800 °C in N$_2$ and O$_2$, respectively. For TDTR measurements, a layer of ~80-nm-thick Al was deposited by DC (direct current) sputtering on the surface of all samples. The thermal conductivity of the Al film is ~170 W/m-K by measuring its electrical conductivity and applying the Wiedemann-Franz Law.

**TDTR Measurements**. The TDTR system used in this work is a two-tint pump-probe system.[16] A mode-locked femtosecond Ti-sapphire laser is split into a pump beam and a probe beam. The pump beam modulated by an electro-optic modulator heats the sample surface periodically while the probe beam detects the temperature variation of the sample surface via thermoreflectance.[13] The signal picked up by a photodetector and a lock-in amplifier is fitted with an analytical heat transfer solution of the sample structure to infer unknown thermal properties.[17] A 10X objective is used in this work with beam sizes of 5.5 μm in radius. The local thicknesses of Al and β-Ga$_2$O$_3$ films were determined by the picosecond acoustic technique.[18] The heat capacity data are from literature.[19,20]

**Supporting Information:** The supporting information includes the data processing of correcting the detector voltage with a linear laser power decaying with time.

**Competing interests:** The authors claim no competing financial interests.

**ACKNOWLEDGEMENT**: Z.C. and S. G. would like to acknowledge the financial support from U.S. Office of Naval Research under a MURI program (Grant No. N00014-18-1-2429). F.M., and T.S. would like to acknowledge the financial support from Japan JSPS KAKENHI Grant Number 19K15298 and the Precise Measurement Technology Promotion Foundation. T. Y., W. X., and X. O. was supported by National Natural Science Foundation of China (No. 61874128, 61851406, and 11705262).

# REFERENCES

1   Tsao, J. *et al.* Ultrawide-Bandgap Semiconductors: Research Opportunities and Challenges. *Advanced Electronic Materials* 4, 1600501 (2018).

2   Cheng, Z. *et al.* Thermal conductance across harmonic-matched epitaxial Al-sapphire heterointerfaces. *Communications Physics* 3, 1-8 (2020).

3   Cheng, Z., Mu, F., Yates, L., Suga, T. & Graham, S. Interfacial Thermal Conductance across Room-Temperature-Bonded GaN/Diamond Interfaces for GaN-on-Diamond Devices. *ACS Applied Materials & Interfaces* 12, 8376-8384 (2020).

4   Gaskins, J. T. *et al.* Thermal Boundary Conductance Across Heteroepitaxial ZnO/GaN Interfaces: Assessment of the Phonon Gas Model. *Nano letters* 18, 7469-7477 (2018).

5   Huxtable, S., Cahill, D. G., Fauconnier, V., White, J. O. & Zhao, J.-C. Thermal conductivity imaging at micrometre-scale resolution for combinatorial studies of materials. *Nature Materials* 3, 298-301 (2004).

6   Sood, A. *et al.* Direct visualization of thermal conductivity suppression due to enhanced phonon scattering near individual grain boundaries. *Nano letters* 18, 3466-3472 (2018).

7   Sood, A. *et al.* An electrochemical thermal transistor. *Nature communications* 9, 1-9 (2018).

8   Zhao, J.-C., Zheng, X. & Cahill, D. G. Thermal conductivity mapping of the Ni–Al system and the beta-NiAl phase in the Ni–Al–Cr system. *Scripta Materialia* 66, 935-938 (2012).

9   Grimm, D. *et al.* Thermal conductivity of mechanically joined semiconducting/metal nanomembrane superlattices. *Nano letters* 14, 2387-2393 (2014).

10  Brown, D. B. *et al.* Spatial Mapping of Thermal Boundary Conductance at Metal–Molybdenum Diselenide Interfaces. *ACS Applied Materials & Interfaces* 11, 14418-14426 (2019).

11  Yang, J., Maragliano, C. & Schmidt, A. J. Thermal property microscopy with frequency domain thermoreflectance. *Review of Scientific Instruments* 84, 104904 (2013).

12  Olson, D. H., Braun, J. L. & Hopkins, P. E. Spatially resolved thermoreflectance techniques for thermal conductivity measurements from the nanoscale to the mesoscale. *Journal of Applied Physics* 126, 150901 (2019).

13  Cahill, D. G. Analysis of heat flow in layered structures for time-domain thermoreflectance. *Review of scientific instruments* 75, 5119-5122 (2004).

14  Cheng, Z. *et al.* Thermal Transport across Ion-cut Monocrystalline β-Ga2O3 Thin Films and Bonded β-Ga2O3-SiC Interfaces. *ACS Applied Materials & Interfaces* (2020).

15  Mu, F., Wang, Y., He, R. & Suga, T. Direct wafer bonding of GaN-SiC for high power GaN-on-SiC devices. *Materialia* 3, 12-14 (2018).

16  Kang, K., Koh, Y. K., Chiritescu, C., Zheng, X. & Cahill, D. G. Two-tint pump-probe measurements using a femtosecond laser oscillator and sharp-edged optical filters. *Review of Scientific Instruments* 79, 114901 (2008).

17  Cheng, Z. *et al.* Probing Growth-Induced Anisotropic Thermal Transport in High-Quality CVD Diamond Membranes by Multi-frequency and Multi-spot-size Time-Domain Thermoreflectance. *ACS applied materials & interfaces* (2018).

18  Cheng, Z. *et al.* Thermal conductance across β-Ga2O3-diamond van der Waals heterogeneous interfaces. *APL Materials* 7, 031118 (2019).

19  Jiang, P., Qian, X., Li, X. & Yang, R. Three-dimensional anisotropic thermal conductivity tensor of single crystalline β-Ga2O3. *Applied Physics Letters* 113, 232105 (2018).

20  Zheng, Q. *et al.* Thermal conductivity of GaN, GaN 71, and SiC from 150 K to 850 K. *Physical Review Materials* 3, 014601 (2019).